\documentclass[manuscript,screen]{acmart}

\AtBeginDocument{%
  }

\copyrightyear{2023} 
\acmYear{2023} 
\setcopyright{acmcopyright}\acmConference[LAK 2023]{LAK23: 13th International Learning Analytics and Knowledge Conference}{March 13--17, 2023}{Arlington, TX, USA}
\acmBooktitle{LAK23: 13th International Learning Analytics and Knowledge Conference (LAK 2023), March 13--17, 2023, Arlington, TX, USA}
\acmPrice{15.00}
\acmDOI{10.1145/3576050.3576149}
\acmISBN{978-1-4503-9865-7/23/03}
\usepackage[shortlabels]{enumitem}

\begin{document}

\title[Demographic and Behavioral Oversampling for Fair Student Success Modeling]
{Protected Attributes Tell Us Who, Behavior Tells Us How: A Comparison of Demographic and Behavioral Oversampling for Fair Student Success Modeling}

\author{Jade Maï Cock}
\email{jade.cock@epfl.ch}
\authornotemark[1]
\affiliation{%
  \institution{EPFL}
  \streetaddress{EPFL IC IINFCOM ML4ED
INF Building – Station 14}
  \city{Lausanne}
  \state{Vaud}
  \country{Switzerland}
  \postcode{CH-1015}
}

 \author{Muhammad Bilal}
\affiliation{%
 \institution{Pakistan Institute of Engineering and Applied Sciences}
 \streetaddress{Department of Computer & Information Sciences, Lehtrar Road}
 \city{Nilore}
 \state{Islamabad}
 \country{Pakistan}
 \postcode{45650}}
 \email{muhammadbilal@pieas.edu.pk}

\author{Richard Lee Davis}
\affiliation{%
  \institution{EPFL}
  \streetaddress{RLC D1 740 (Rolex Learning Center)
Station 20}
  \city{Lausanne}
  \state{Vaud}
  \country{Switzerland}
  \postcode{CH-1015}}
\email{richard.davis@epfl.ch}

\author{Mirko Marras}
\affiliation{%
 \institution{University of Cagliari}
 \streetaddress{Department of Mathematics and Computer Science}
 \city{Cagliari}
 \state{Sardinia}
 \country{Italy}
 \postcode{09124}}

\author{Tanja Käser}
\affiliation{
  \institution{EPFL}
  \streetaddress{EPFL IC IINFCOM ML4ED INF Building – Station 14}
  \city{Lausanne}
  \state{Vaud}
  \country{Switzerland}
  \postcode{CH-1015}}
  \email{tanja.kaser@epfl.ch}

\renewcommand{\shortauthors}{Cock et al.}

\begin{abstract}
Algorithms deployed in education can shape the learning experience and success of a student. It is therefore important to understand whether and how such algorithms might create inequalities or amplify existing biases. In this paper, we analyze the fairness of models which use behavioral data to identify at-risk students and suggest two novel pre-processing approaches for bias mitigation. Based on the concept of intersectionality, the first approach involves intelligent oversampling on combinations of demographic attributes. The second approach does not require any knowledge of demographic attributes and is based on the assumption that such attributes are a (noisy) proxy for student behavior. We hence propose to directly oversample different types of behaviors identified in a cluster analysis. We evaluate our approaches on data from (i) an open-ended learning environment and (ii) a flipped classroom course. Our results show that both approaches can mitigate model bias. Directly oversampling on behavior is a valuable alternative, when demographic metadata is not available. Source code and extended results are provided in \href{https://github.com/epfl-ml4ed/behavioral-oversampling}{https://github.com/epfl-ml4ed/behavioral-oversampling}.
\end{abstract}



\maketitle

\section{Introduction}
Algorithms deployed in an educational context have the power to shape a student's future both on a small and large scale. These algorithms can for example help students understand a scientific concept \cite{hensberry2015effective}, provide assistance for teachers in grading assignments \cite{kakkonen2006applying}, or support educators in identifying struggling students early \cite{swamy2022meta}.
In a world where education is already unfair or biased towards certain communities \cite{caldas1997effect, tomul2012socioeconomic,kong2020reducing, robnett2016gender}, it is particularly important to understand whether and how the algorithms we deploy might perpetuate or even amplify existing inequalities \cite{baker2021algorithmic, kizilcec2020algorithmic}. In this paper, we therefore analyze the (un)fairness of models that adopt behavioral data to identify at-risk students and explore methods for remediating the biases that led to unfairness.

Gender, socio-economic background, cultural background, and country have all been found to be linked to individuals' learning strategies and expectations towards education (e.g., \cite{astleitner2005there, meyer1995gender}).
parents' socio-economic and cultural background are related to school completion and academic performance \cite{kilicc2012comparison}, and country is linked to students' metacognition of academic achievement \cite{kilicc2012comparison}. 
Demographic differences such as these can be problematic for machine learning algorithms. 
In the case where datasets are imbalanced in favor of one demographic attribute (e.g., gender), the model may over-focus on the idiosyncratic characteristics of the over-represented group, leading to better predictions for that population while also making worse predictions for other under-represented groups.
This is particularly problematic when demographic groups who are under-represented in the data are also under-represented in society (and \emph{protected} for instance by anti-discrimination legislation), as a model's worse performance for these groups might amplify existing societal biases.
An example of one such model is the algorithm developed to predict students' prospective GCSE grades in the UK during the pandemic \cite{smith2020algorithmic}. Around $40\%$ of the students were downgraded compared to their teachers' predictions, and a majority of them were disadvantaged students. Yet, wealthier and more privileged students tended to benefit from that model.
Due to these risks, investigations into algorithmic fairness have permeated nearly all areas of the artificial intelligence field \cite{mehrabi2021survey, pessach2020algorithmic}. This research aims to formalize the notions of algorithmic unfairness, to develop methods of identifying when algorithms are unfair, and to develop bias mitigation methods in pre-, in-, and post-processing.

In this paper, we investigate the fairness of models for the early detection of at-risk students using behavioral data. In recent years, a number of biases have been uncovered in student success prediction models \cite{mehrabi2021survey, kizilcec2020algorithmic, baker2021algorithmic}, and a variety of mitigation methods have been put forward \cite{hu2020towards, yanfair2020, yu2020towards}. 
Pre-processing approaches that oversample based on both demographic attributes and outcomes have led to improved model fairness \cite{ranvcic2021investigating, imbalanced-learn}.
These approaches have targeted a representation bias, where a part of the population is under-represented \cite{anderson2019assessing}. 
The common approach is to identify demographic imbalances in the data and to oversample the data in these categories to create a more balanced training dataset. For example, \cite{ranvcic2021investigating} found that oversampling on gender and race reduced disparate impact, while \cite{salazar2021fawos} showed that re-balancing demographic groups by taking into account ground truth labels reduced the disparate impact.

While these approaches are promising, they are also general.
We have identified two novel pre-processing approaches that are better tailored to the educational context as well as to the data and modeling approaches used in the at-risk detectors.
The first, \textit{guided demographic oversampling}, involves oversampling on combinations of demographic attributes instead of simply oversampling on a single category.
The second, \textit{behavioral oversampling}, starts by identifying behavioral learning profiles, and then upsamples the profiles which are under-represented in the data.

Instead of oversampling on a single attribute such as gender, race, or geographic location, \textit{guided demographic oversampling} involves first splitting the data into smaller buckets which represent combinations of attributes (e.g., female-high-SES, male-high-SES, female-low-SES, male-low-SES) and then upsampling under-represented groups in order to re-balance representation among these combined categories.
This approach is inspired by the concept of intersectionality originally introduced by \cite{crenshaw2018demarginalizing} to describe how the forms of discrimination faced by Black women cannot be reduced to those experienced by either Black people or women.
Applied more broadly, intersectionality describes how unique forms of bias emerge from the interaction of intersecting identities.
Following this line of reasoning, when attempting to mitigate algorithmic bias there may be value in oversampling subgroups containing students with intersecting identities, as opposed to more standard techniques which only oversample on individual attributes.

The second approach that we introduce is \textit{behavioral oversampling}.
It sidesteps issues related to demographic oversampling by focusing instead on attaining a balanced representation of student behaviors in the data.
The models used for detecting at-risk students are trained on time-based sequences of behavioral data and learn to identify patterns of behavior that are associated with a higher probability of getting a low grade, failing, or dropping a course.
When demographic oversampling is successful at mitigating bias, this is likely because certain under-represented demographics exhibit behaviors that can be associated with various outcomes.
Through oversampling, we force the model to pay more attention to these behaviors allowing it to make more accurate and less biased predictions for these under-represented groups.
While demographics may serve as a reliable proxy for under-represented behaviors in many cases, this relationship might not always hold \cite{lee2022effects}. In these situations, demographic oversampling will have little effect on mitigating biases, because under-represented behaviors are not confined to a single demographic group. 

Behavioral oversampling has the potential to bypass this problem. 
Rather than using demographic attributes as proxies for learning strategies, we propose to oversample directly on learning behaviors. This approach has the additional advantage that it works in the absence of access to demographic attributes which are not always easy to obtain. Our approach extends \cite{yanfair2020}'s work who used clusters and borderline SMOTE to re-balance datasets 
through a pipeline which relies on time series data points and random oversampling on the full data to balance behavioral educational datasets.
We evaluate both proposed approaches on models which use behavioral data to identify at-risk students in two vastly different learning contexts: a flipped classroom course and an open-ended learning environment. Specifically, we aim to answer three research questions: 1) What biases do we find in early prediction models and how are they related to demographic under-representation? 2) To what extent can oversampling on combined attributes (\textit{guided attribute oversampling}) reduce unfairness of at-risk detectors, especially in comparison to single-attribute oversampling? 3) Does oversampling on behavioral clusters (\textit{behavioral oversampling}) help to reduce unfairness, and how does it compare to demographic oversampling?
Our results show that biases in early prediction models are not necessarily related to an under-representation of a demographic attribute and that combinations of attributes constitute a more reliable proxy of learning behavior. Moreover, our findings show that biases can be mitigated using solely behavioral data.

\section{Data and Methods}\label{section:methods}
\begin{figure}[t]
  \centering
  \includegraphics[width=1.0\linewidth]{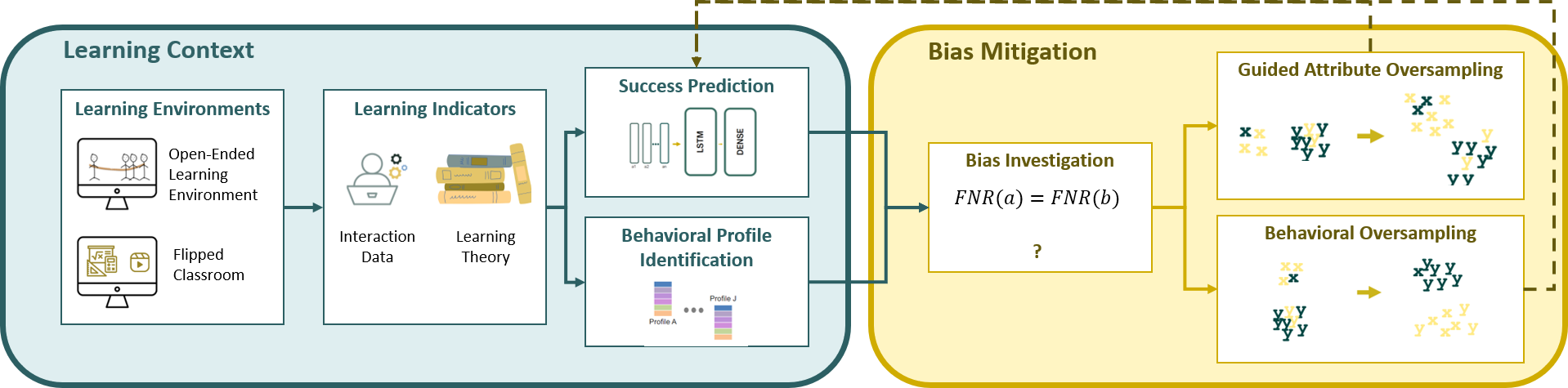}
  \caption{Bias mitigation framework: we extract interaction data from a flipped classroom and open-ended learning environment (OELE) context to build learning indicators. These indicators serve as an input for the at-risk detectors as well as for the clustering algorithm. In the bias mitigation part, we analyze the bias of the success predictions in detail and perform a guided attribute oversampling based on imbalances detected in the data set. Furthermore, we also oversample based on the profiles identified in the clustering part.}
  \Description{Bias mitigation framework: we extract interaction data from a flipped classroom and open-ended learning environment (OELE) context to build learning indicators. These indicators taken from learning theories serve as an input for the at-risk detectors as well as for the clustering algorithm. In the bias mitigation part, we analyze the bias of the success predictions in detail and perform a guided attribute oversampling based on imbalances detected in the data set. Furthermore, we also oversample based on the profiles identified in the clustering part.}
  \label{figure:pipeline}
\end{figure}

Our goal in this paper is to study 
(i) the extent to which success prediction models might lead to disparate outcomes across demographic groups 
and (ii) how such disparities can be reduced by acting on the representation of groups in the training data via demographic and behavioral oversampling.
Following the pipeline in Figure \ref{figure:pipeline}, we first collected students' data under two different instructional strategies namely flipped classroom and open-ended exploration. 
For each instructional strategy, we then extracted a range of behavioral indicators found relevant in prior learning analytics work.
With these indicators, we created student success models and identified relevant clusters of students according to their behavior in the activity. 
Subsequently, we performed an exploratory analysis to assess disparities according to demographic groups and behavioral groups created based on the identified clusters.
We finally investigated the extent to which different combinations of demographic and behavioral oversampling techniques can mitigate the disparities.

\vspace{-1mm}  \subsection{Modeling Student Learning} 
Student success models are becoming popular in education for a variety of tasks such as adaptive intervention and learning understanding. 
Indeed, flipped teaching and open-ended exploration are two relevant instructional strategies that can adopt student success models for the above purposes. 
The next subsections will describe the real-world learning contexts considered in our study where teaching and learning were based on the above instructional strategies. 

\vspace{1mm} \noindent \textbf{Flipped Classroom Context}.
Flipped classroom is a course format where students are required to study and learn on top of a certain amount of material as a preparation for class. Usually, course material is shared on a learning platform.
Students then come to class and engage in learning activities in small groups with and without the instructor. 

\vspace{1mm} \noindent \textit{Tracking Student Learning}.
Our analysis was based on pre-class activities pertaining to a compulsory Linear Algebra (LA) undergraduate course (Table \ref{tab:envs}, first row) delivered in a flipped classroom setting by a large European university. 
Pre-class activities were performed on an instance of the Open edX platform where the instructor asked students to watch videos and complete quizzes of an already existing Linear Algebra MOOC.  
Students had access to pre-recorded videos and they had the opportunity to test their competencies through short quizzes after a sequence of videos. 
Furthermore, they could read and download slides or any additional material the instructor provided for the topic. 

The log entries collected by the platform were tuples reporting the user, the activity, and the timestamp (e.g., user: 10, activity: play video 32, timestamp: 05-03-2018 12:06:01). 
The data set was collected from two consecutive flipped editions of the course, taught by the same lecturer and with a flipped duration of 10 weeks. 
Success of students was modeled using the grade of the final exam. Specifically, we aimed at predicting students' intervention need and therefore used the label $1$ for students failing the exam (and thus needing intervention) and the label $0$ for students who passed the exam (and consequently, there was no intervention need).
We also had access to students' self-reported gender and the country of origin of their high school diploma as this information has been shown to correlate with prior knowledge by previous work \cite{hardebolle2022gender}. The study was approved by the institutional ethics review board (HREC 058-2020/10.09.2020).

\vspace{1mm} \noindent \textit{Learning Indicators}. 
A large body of work has focused on engineering learning indicators for success prediction in online or blended settings. In a recent meta-analysis, \cite{marras2021can} built a combined set of learning indicators retrieved from relevant previous work and systematically assessed their effectiveness for success prediction across different online and blended courses. The learning indicators that they found most effective are related to students' self-regulated learning behavior. Self-regulated learning (SRL) characterizes the ability of a student to be responsible of their own learning and has been widely investigated in online learning settings (e.g., \cite{cho2013self,sher2020analyzing}). Based on a meta-analysis on SRL for online higher education \cite{broadbent2015self}, we categorize the learning indicators into the following three dimensions (that are associated with academic achievement): time management (ability to plan study time and tasks), effort regulation (persistence in learning), and metacognition (awareness and control of thoughts). Learning indicators related to time management measure the distribution of students' effort throughout the entire course (\textbf{Consistency}), the extent to which students regularly reserve time to study (\textbf{Regularity}), and the students' tendency to follow the course (\textbf{Proactivity}). Learning indicators categorized into effort regulation measure students' engagement and persistence during the course (\textbf{Effort}). Finally, learning indicators belonging to the metacognition dimension assess students' self-assessment abilities (\textbf{Assignment}) as well as how they control the learning flow within a certain content (\textbf{Control}). The self-regulated learning indicators identified as important by \cite{marras2021can} for the context of flipped classrooms have also been shown to be predictive for a large set of MOOCs \cite{swamy2022meta}. We hence used the suggested indicators of SRL behavior as a basis for our success prediction and profile identification.
Specifically, we represented each student's interaction sequence by a time series of $14$ data points (one for each week), where each data point was represented by an $82$-dimensional vector, and each of those $82$ cells attempted to capture in different ways one of the six SRL aspects described above.

\begin{table*}[!t]
\caption{\textit{Learning contexts}. Detailed information about the flipped and open-ended environments targeted by our study.}
\label{tab:envs}
\vspace{-2mm}
\small
\resizebox{\textwidth}{!}{
\begin{tabular}{lllrrllrrr}
\toprule
\textbf{Course Title} & \textbf{ID} & \textbf{Field} &  \textbf{Setting} & \multicolumn{1}{r}{\textbf{\begin{tabular}[c]{@{}c@{}} Students\end{tabular}}} & \textbf{Level} & \textbf{Language} & \multicolumn{1}{r}{\textbf{\begin{tabular}[c]{@{}c@{}} Duration\end{tabular}}} &  \multicolumn{1}{r}{\textbf{\begin{tabular}[c]{@{}c@{}} Failing Rate \end{tabular}}} & \multicolumn{1}{r}{\textbf{\begin{tabular}[c]{@{}c@{}} Demographics$^1$\end{tabular}}} \\
\midrule
Linear Algebra & \texttt{LA} & Mathematics & Flipped course & 191 & BSc & English & 10 weeks & 42\% & F: 35\%, M: 65\%; CO: 49\%, CF:47\%\\ 
 Tuglet & \texttt{Tuglet} & Science & Educational game & 264 & Middle School & English & $\leq$ 15. min & 47\% & F: 48\% M: 51\%; H: 52\%, M:48\%\\
\bottomrule
\end{tabular}}
\scriptsize{$^1$ \textbf{Demographics} GF : Female ; GM : Male; CO : Original Origin; CF : Foreign Country, H: district with high socio-economical status (SES), M: district with medium SES.}
\end{table*}

\vspace{1mm} \noindent \textit{Behavioral Profile Identification}.
\label{sec:flipped_clustering}
To identify behavioral profiles, we relied on \cite{mejia2022identifying}'s work, which used the above learning indicators to cluster students for the LA course into groups with similar SRL behavior. They proposed a multi-step clustering pipeline where they first clustered students separately for each SRL dimension. Then, they integrated the resulting behavioral patterns into multi-dimensional profiles. They obtained five distinct SRL profiles for the LA course and showed that these profiles were associated with learning outcomes. Students in profile \textbf{'E'} worked consistently during the semester, reserved regular time slots for learning, were in-sync with the course level and exhibited a high level of metacognition. Students in profile \textbf{'D'} showed a very similar SRL behavior to profile \textbf{'E'}, with the only difference being that they were delayed in submissions with respect to the course schedule. Profile \textbf{'C'} differed from profile \textbf{'E'} also in only one dimension; students in this profile were much more engaged (higher effort). Students in profile \textbf{'B'} had a mixed SRL behavior. While they showed a high level of metacognition, they struggled with time management. Finally, students in profile \textbf{'A'} struggled in all dimensions. They were not very engaged, did not work consistently and regularly, were delayed in their coursework, and showed a low level of metacognition. All profiles are described in \cite{mejia2022identifying}.

\vspace{1mm} \noindent \textit{Success Prediction}.
\label{sec:flipped_prediction}
Recent work has suggested a series of models for student success prediction in online and flipped setting. Since we were interested in predicting intervention needs (rather than success), in the following we will denote our model as an \textit{at-risk detector}. For our study, we relied on the neural architecture based on \emph{Bidirectional LSTMs} proposed by \cite{swamy2022meta}. The architecture is composed of two simple \emph{BiLSTM} layers of size $8$ (using a lookback of $3$) and a \emph{Dense} layer (with Sigmoid activation) having a hidden size of $1$. The indicated hyperparameters have been found using a nested student-stratified (i.e. dividing the folds by students) 10-fold cross validation. 
\vspace{1mm} \noindent \textbf{Open-Ended Exploration Context}.
Over the last decade, there has been an increase in the use of open-ended learning environments (OELEs) such as educational games or simulations. 
However, many students struggle in efficiently doing problem-solving and inquiry in these environments \cite{kaser2017modeling,cock2021early}. 
Modeling students’ learning as they try to benefit from OELEs may be a useful next step in educators' abilities to support students’ development of exploration strategies.
\vspace{1mm} \noindent \textit{Tracking Student Learning}.
Our study was based on TugLet \cite{DBLP:conf/lak/KaserHS17}, an interactive game for assessing students' inquiry strategies. The game revolves around a tug-of-war between two teams composed of figures of different strengths (large: $3$, medium: $2$, small: $1$). Players can choose between two different modes: in \textit{Explore} mode, they can simulate different compositions of teams and observe the outcome of the tug-of-war. In \textit{Challenge} mode, they have to predict the outcome of a given tug-of-war and receive right/wrong feedback. If they make a mistake, they are sent back to the\textit{Explore} mode. They are, however, free to return to \textit{Challenge} mode at any point in time. The game is over when the player manages to predict the outcome of eight tug-of-wars (of increasing difficulty) correctly in a row. After that, a post-test assesses whether the players have learned the relationships between the strengths of the different figures. We used students' posttest score to model success in the game. We aimed at predicting intervention need and therefore used the label $1$ ($0$) for students with a low (high) posttest score and thus needing (not needing) intervention.
TugLet was equipped with a logging system which recorded 
all the trials made in \textit{Explore} and \textit{Challenge} along with their correctness (for \textit{Challenge} mode only) and a time stamp. The data used in this study was collected in a classroom experiment in two different North American middle schools with a total of $365$ students. The first school had a medium socio-economic status (SES) while the latter had a medium to high SES. We also collected the gender of the students. The study was approved by the institutional ethics review board (HREC 060-2020/04.09.2020).

\vspace{1mm} \noindent \textit{Learning Indicators}.
\label{sec:tuglet_features}
The open-ended nature of the tasks in OELEs makes representing and predicting student success a challenge. 
We relied on prior work to build the learning indicators of TugLet. \cite{DBLP:conf/lak/KaserHS17} modeled students' exploration strategies in combination with the correctness of their answers to predict student success in TugLet. Specifically, they formalized the principles that the students could learn through the game using a set of 'rules'. These rules cover the logic of the whole game. Example rules are: 'a large figure (3) is stronger than a medium figure (2)' or 'a large figure (3) is as strong as a medium (2) +  a small figure (1)'. Each trial was then characterized by attaching the set of rules needed to predict the winning side. More importantly, the quality of the trial can be determined by the size of the rule set. For example, a trial with a large rule set is considered as weak, as a large rule set indicates a complex team composition, which does not allow drawing any conclusions about the relationship between the strengths of the different figures. A trial with only one associated rule is considered as strong its simplicity means that the relationships can be directly derived. \cite{DBLP:conf/lak/KaserHS17} demonstrated that exploration quality (measured by the size of the rule set) was related to achievement in the posttest.
Given these learning indicators, we represented each student in our dataset by a sequence of $n$ points where $n$ denotes the total number of trials of that student. Each of these data points captured the game mode, rule set associated with the trial, team compositions, winning team, time taken to simulate/predict, and trial correctness (if in \textit{Challenge} mode).

\vspace{1mm} \noindent \textit{Behavioral Profile Identification}. In order to identify the behavioral profiles for TugLet, we derived on the clusters identified by \cite{kaser2020modeling}. They used the rules and exploration quality metric described above to build three time series features for each student: the cumulative number of Explore trials up to a specific point in the game, the cumulative number \textit{Challenge} trials up to a specific point in the game, and the cumulative number of Explore trials classified as strong up to a certain point in the game. They used k-means clustering on these indicators and found six different profiles characterized by exploration behavior and efficiency in passing the game. The first three clusters all passed the game fast. The \textbf{systematic explorers} utilized inquiry processes in order to uncover the principles of the game. The \textbf{explorers} also used inquiry, but were less systematic. The \textbf{non explorers} did not simulate a single team composition, but tried to pass the game via trial-and-error in \textit{Challenge} mode. The second three clusters took a long time to pass the game. The \textbf{slow explorers} simulated many team compositions without any system. The \textbf{slow passers} did not use simulation at all. The \textbf{mixed explorers} simulated some team compositions. The interpretation of such profiles can be found in \cite{kaser2020modeling}.

\vspace{1mm} \noindent \textit{Success Prediction}.
Recent work has successfully applied LSTMs to predict student success in an OELE \cite{cock2021early}. We hence adopted LSTMs using students' interaction sequences represented as a series of trials as an input. Each trial is described by the game mode, the rule set associated with the trial, the team compositions, the winning team, the time taken to simulate/predict, and the trial correctness (if in \textit{Challenge} mode). The architecture is composed of one LSTM layer of size $32$ and a \emph{Dense} layer (with Sigmoid activation) having a hidden size of $1$. We used a dropout of $0.02$. The hyperparameters have been found using a nested student-stratified (i.e. dividing the folds by students) 10-fold cross validation. Again, we will also denote this model as an \textit{at-risk detector} as we are predicting intervention needs (rather than success). 

\vspace{-2mm}
\subsection{Unfairness Diagnosis and Treatment}
\label{sec:biasmitigation}
In the case where the predictors are found to provide biased predictions, we attempt to mitigate the resulting unfairness by re-balancing the dataset using random oversampling with replacement. In the rest of this section, we define and formalize our notion of \textit{biased predictions}. Then, we motivate and detail the process by which we create the groups we oversample on which is the main contribution of this work. Finally, we lay out the actual oversampling process.

\vspace{1mm} \noindent \textbf{Unfairness Assessment}. 
Extensive work has been conducted on designing metrics with the purpose of capturing the different unfair behavior that can occur in machine learning scenarios.
Unfortunately, some of the fairness principles have been translated into metrics which cannot be simultaneously satisfied \cite{kleinberg2018inherent}.
Consequently, to identify the fairness principles we want to prioritize, it is important to identify where the biases may occur and estimate their short- and long-term impact on students.
In this study, we aim to support students at risk of failure by suggesting targeted remedial interventions before the final evaluation of the topic at hand.
Therefore, it is essential to identify all students at risk of failing by reducing false negative rates (FNR) taking into account that some students might be offered unnecessary support (the false positives).
Additionally, we aim to mitigate biases when identifying at-risk students and consequently we focus on reducing the differences in FNR between the different classes determined by a certain attribute $o$.

\vspace{1mm} \noindent \textbf{Unfairness Treatment}. 
As previously discussed, demographic attributes such as gender and the country of education are typically related to students' learning behaviors.
These differences across demographic characteristics can be problematic when models are being built and trained on behavioral features.
In the case where the data sets are imbalanced towards one demographic attribute, the model will tend to over-focus on the idiosyncratic behavioral characteristics of that group,- while ignoring different behaviors exhibited by under-represented groups.
One way of ameliorating this undesirable behavior is to ensure that the dataset used to train the model contains an equal representation of the different groups in the population.
This is typically achieved through oversampling the under-represented groups in order to achieve a more balanced training data set.
Prior work on bias detection and mitigation in early detection models has found success using generic oversampling techniques such as oversampling under-represented demographic groups \cite{imbalanced-learn} and oversampling on a combination of demographics and outcomes \cite{salazar2021fawos}.
We hypothesize that these general methods have weaknesses that are specific to the domain of early detection of student failure using behavioral data.
To address these weaknesses, we propose two new methods for oversampling that are better tailored to the domain: \textbf{guided demographic oversampling} and \textbf{behavioral oversampling}. 

\vspace{1mm} \noindent \textit{Guided Demographic Oversampling}.
Common approaches to oversampling on demographic attributes start by identifying algorithmic biases that affect under-represented groups and then proceed by oversampling data points from these groups when constructing the training dataset \cite{ranvcic2021investigating, salazar2021fawos}.
There are two issues with this approach: a practical one and a theoretical one.
The practical issue is that this approach cannot be used when the data set is already balanced with respect to demographic attributes.
The theoretical issue is that oversampling on a single demographic attribute does not take intersectional biases into account.
Intersectional biases are those that affect people with specific combinations of demographic attributes such as women from a specific country or men from low socio-economic backgrounds.

Guided demographic oversampling is designed to address both of these issues.
Formally, let $o$ be an attribute on which an algorithm is biased.
Then, we assume that the different classes of $o$ are either imbalanced, or that the bias is caused by an interaction of $o$ with another attribute.
Concretely, if an at-risk detector works better for one gender than for others, the likely cause is either that the number of individuals varies across genders or that the number of individuals varies across groups which are defined by their gender and another attribute.
To investigate these bias-imbalances systematically, we first investigate whether there are imbalances in standalone attributes.
Standalone attributes with imbalanced classes are added to the set $\mathcal{O}$.
Then, if an attribute on which the detectors are biased is naturally balanced, we investigate whether combinations of attributes create imbalanced groups. When that is the case, we add that combination of attributes to $\mathcal{O}$. By doing so, $\mathcal{O}$ will contain all the combinations of attributes that lead to the creation of imbalanced groups. For each of the attributes or combination of attributes $a$ in $\mathcal{O}$, we oversample the under-represented groups to create a more balanced data set, and then continue training the early detection models as usual.

\vspace{1mm} \noindent \textit{Behavioral Oversampling}. 
Though we hypothesize that guided attribute oversampling can have an impact on fairness, there are two issues that this method does not address.
The first issue is a practical one, which is that demographic information is not always available, and that when it is available it sometimes cannot be used for ethical reasons or because of data-protection laws.
A more fundamental issue is that oversampling on demographic information assumes that demographic attributes are proxies for behavioral patterns.
If an early detection model trained on behavioral data exhibits bias against a demographic group that is under-represented in the data, one explanation is that this group exhibits idiosyncratic behaviors that the model is unable to pick up on.
When data from this group is oversampled, this has the effect of amplifying these idiosyncratic behaviors, allowing the model to learn the associations between them and educational outcomes.
However, the assumption that the members of under-represented demographic subgroups all exhibit similar and idiosyncratic behaviors is strong, yet in many situations unlikely to hold.

We introduce behavioral oversampling in response to these issues.
The first step is to identify groups of students which exhibit similar patterns of behavior.
Here, we cluster students in behavior space, though other approaches may be valid as well.
More formally, we use the learning profiles implemented by \cite{kaser2020modeling} for TugLet and the profiles identified by \cite{mejia2022identifying} for flipped classrooms as the behavioral groups we consider for oversampling.
We then investigate the balance of the clustering solution. An imbalance in cluster sizes indicates that some learning behaviors are under-represented in the data and we hence oversample these groups to construct a more balanced training dataset, and then continue training the at-risk detectors.

\vspace{1mm} \noindent \textit{Oversampling Techniques}. 
To re-balance the classes (clusters), we apply different methods which all use random oversampling as implemented in \texttt{imbalanced-learn}~\cite{imbalanced-learn}. These methods are described below:
\begin{itemize}
    \item \textbf{Equal Balancing} re-balances equally each class defined by $o$ (\texttt{sampling\_strategy='all'} in imbalanced-learn). The instances chosen to upsample are randomly selected with replacement \cite{imbalanced-learn, salazar2021fawos}.  
    \item \textbf{Majority Oversampling} upsamples the majority class as defined by $o$ by $50\%$. Because of the small dataset sizes, oversampling on very small categories might introduce noise, while oversampling on the majority may increase the signal in the training set \cite{jeonggets}. This oversampling parameter is used for binary cases to boost the signal in the dataset. For more classes, only oversampling one group has a lesser effect.
    \item \textbf{Cascade Oversampling} upsamples each class defined by $o$ gradually such that any of these classes gets upsampled to the size of the smallest larger class. For example, if class $o_1$ is of size $7$, class $o_2$ is of size $15$ and class $o_3$ is of size $3$, then $o_2$ remains the same as it is the largest one, $o_1$ is upsampled to $15$, and $o_3$ is upsampled to $7$. Similarly to majority oversampling, cascade oversampling amplifies the signal in the dataset, while augmenting the visibility of the under-represented class without introducing too much noise or too many similar individuals in the training. We apply this technique to datasets with binary classes. 
    \item \textbf{Minor Oversampling} upsamples the minority classes to the number of instances present in the majority class. When the minority class contains less than $10$ samples, we upsample the second smallest class of the dataset to avoid the introduction of noise by constantly repeating the same samples \cite{imbalanced-learn}. 
    \item \textbf{Within Oversampling} only applies when oversampling on a combination of attributes. Then, a main attribute is chosen (in our case it is always the clusters) and the other attributes are first rebalanced within their main attribute before rebalancing the main attributes. For example, let \textit{behavior} and \textit{gender} be the attributes by which we want to oversample, let $A$ and $B$ be the existing clusters, and let $t$ and $d$ be the genders people identify with in our dataset. Let the distribution of the data be as followed: $|A\cap t| = 12$, $|A\cap d| = 8$, $|B\cap t| = 6$ and $|B\cap d|= 4$. Then, when oversampling in a \textit{within} manner, we first rebalance gender within the intervention need groups and obtain: $|A \cap t| = 12$, $|A\cap d| = 12$, $|B\cap t| = 6$ and $|B\cap d|=6$. We then use those freshly oversampled groups to once again re-balance the intervention need groups such that $|A| = 24$ and $|B| = 24$.
\end{itemize}
Particularly, we re-balance the training set at each fold according to one of those three methods.
Because we cannot risk developing an algorithm that appears "fair" during training but in fact is not on our test set, we first run all five manners of oversampling before choosing that with the lowest difference of false negative rates across classes. Specifically, we compute the average of 1) the difference in FNR between the two genders, 2) the average in FNR between the two geographical attributes, and then pick the one with the lowest mean. If the FNR rate of the chosen configuration is much higher (15\%) than the original one, we look into the second lowest average and see if the trade off is worth it or not.

\section{Experimental Evaluation} 
\label{sec:results}
We conducted experiments in both the flipped classroom and OELE context to study
(i) the extent to which success prediction model might lead to disparate outcomes across demographic groups and (ii) how such disparities can be reduced via demographic and behavioral oversampling. In the following, we discuss the experimental setup, before detailing out the results of our bias investigation, attribute oversampling, and finally behavioral oversampling.

\subsection{Experimental Setup}
\label{sec:setup}
The predictions of the at-risk students' detectors build the basis of our bias investigation and mitigation. In our experiments, we first trained and tested the original detectors (\textit{baseline}) on their respective data sets. We then also trained and tested the detectors for the \textit{guided attribute oversampling} as well as the \textit{behavioral oversampling}. 

For all three settings, we used a \textit{$10$-fold student stratified cross validation} to evaluate the accuracy and fairness of the model. We did not perform any hyperparameter tuning, but used the hyperparameters optimized for the original detectors (see Section \ref{section:methods}). The stratification was performed on the predicted label ($1$: needs intervention, $0$: no intervention needed). Predictive performance was evaluated using the area under the ROC curve (AUC). We used the AUC as a performance measure as it is robust to class imbalance. Furthermore, as described in Section \ref{sec:biasmitigation}, we analyze the differences in the false negative rate (FNR) as a fairness metric. In our reporting, we exclude categories of attributes which contain less than ten representatives (they were, however, included in the model training and performance evaluation).
In the guided attribute oversampling, we investigated the gender, geographical dimension, and pass-fail label. We used the learning profiles identified through clustering \cite{kaser2020modeling, mejia2022identifying} as a basis for the behavioral oversampling.

\subsection{RQ1: Bias Investigation}
\label{sec:rq1}
In a first analysis, we were interested in analyzing differences in FNR with respect to the investigated attributes and to combinations of these attributes for both learning contexts. Though we focus on false negative rates because we prioritise retrieving all students requiring help, we also investigated false positive rates (FPRs) in our experiments. For conciseness, FPRs are provided in our Github repository ("Reports" folder). Previous work in different learning contexts suggests that model unfairness towards a group with specific demographic attributes is the result of an under-representation (here established with a 15\% difference in representation compared to the majority) of this group in the data set \cite{ranvcic2021investigating, salazar2021fawos}. The TugLet data set is almost balanced in terms of gender (males: $51\%$ , females: $48\%$, see also Table \ref{tab:envs}) and SES (School M: $48\%$ , School H: $52\%$, see also Table \ref{tab:envs}), and we therefore hypothesized that the at-risk detector for TugLet would be fair with respect to these attributes (\textit{H1-1}). In the flipped classroom context, we observed an under-representation of females (males: $65\%$ , females: $35\%$, see also Table \ref{tab:envs}), while the data set was balanced in terms of country of diploma (Country CF: $47\%$ , Country CO: $49\%$, see also Table \ref{tab:envs}). We therefore hypothesized that the at-risk detector for the flipped context would be fair with respect to the country, but might disadvantage female students due to their under-representation in the data set (\textit{H1-2}).

Table \ref{tab:baseline} illustrates the overall accuracy in terms of AUC as well as the FNR per demographic attribute for the \textit{baseline} at-risk detector. For TugLet, we
observe that the detector exhibits a higher FNR for students from School H ($FNR: 0.73$) than for students from School M ($FNR: 0.57$). Students from School H are therefore less likely to receive intervention when needed. Similarly, the at-risk detector disadvantages males ($FNR: 0.70$) who are less likely to be identified at-risk when struggling than females ($FNR: 0.53$).
For the flipped classroom data set, we also observe differences in FNR across demographic attributes. The detector is more likely to fail identifying students needing intervention when they have a diploma from Country CF ($FNR_{Country CF}: 0.58$, $FNR_{Country CO}: 0.42$). In terms of gender, females were disadvantaged by the detector ($FNR_{male}: 0.43$, $FNR_{female}: 0.56$).

\vspace{1mm} \noindent \textit{To summarize, we observed differences in FNR for both learning contexts. While the TugLet data set was balanced in terms of demographic attributes, the at-risk detector still disadvantaged students with specific attributes and we can therefore reject \textbf{H1-1}. In the flipped context, the under-represented group (females) is indeed disadvantaged, but the detector also shows bias with respect to the country. We therefore partially accept \textbf{H1-2}}.

\begin{table*}[t]
\centering
\caption{Baseline performance and fairness results in terms of AUC and FNR. It can be observed that there are differences in FNR across demographic attributes for both learning contexts.}
\label{tab:baseline}
\resizebox{\textwidth}{!}{
\begin{tabular}{c|c||cccccc}
\toprule
 & \textbf{Accuracy (AUC)} & \multicolumn{6}{c}{\textbf{Fairness (FNR)}} \\ 
 & Overall & Females & Males & School A & School B & Country 1 & Country 2 \\
 \midrule
 TugLet & $0.68\pm 0.05$ & $0.53\pm 0.14$ & $0.70\pm 0.19$ & $0.57\pm 0.13$ & $0.73\pm 0.25$ & & \\
 Flipped Classroom & $0.63\pm 0.13$ & $0.56\pm 0.38$ & $0.43\pm 0.26$ & & & $0.58\pm 0.24$ & $0.42\pm 0.29$\\
\bottomrule
\end{tabular}
}
\end{table*}

\subsection{RQ2: Attribute Oversampling}
\label{sec:rq2}
In a second analysis, we investigated the bias of the detectors in both learning contexts in more depth, with the goal to \textit{smartly} upsample the data sets based on observed interactions between combinations of attributes.

\vspace{1mm} \noindent \textit{TugLet}. In our first analysis, we found biases in the detector even though the data set was balanced with respect to gender and SES. We therefore performed more detailed analyses of the interaction effects between combinations of demographic attributes and intervention needs (ground truth labels). We observed that the dataset was also balanced with respect to combined attributes, i.e. the combination of gender and SES of school district (males-School M: $26\%$, males-School H: $25\%$, females-School M: $22\%$, females-School H: $26\%$). Finally, also the ground truth labels were almost balanced (label $1$: $53\%$, label $0$: $47\%$). We did, however, find an interaction between the intervention need and demographic attributes. $58\%$ of the students needing intervention came from school M (ground truth label $1$). There was also a smaller imbalance of intervention needs with respect to gender; female students tended to need intervention more often the male students ($45\%$ of the students needing intervention were males). Overall, these led to an imbalance in intervention need also across combined attributes: from the students needing intervention, $15\%$ were females from school M, $24\%$ were males from school M, $30\%$ were females from school H, and $30\%$ males from school H. We therefore hypothesized that the bias in the detector stemmed from the interaction effect between the gender, SES, and intervention need, and that we could therefore mitigate this bias by upsampling on the full combination of attributes (\textbf{\textit{H2-1}}).

\begin{figure}[t]
  \centering
  \includegraphics[width=\linewidth]{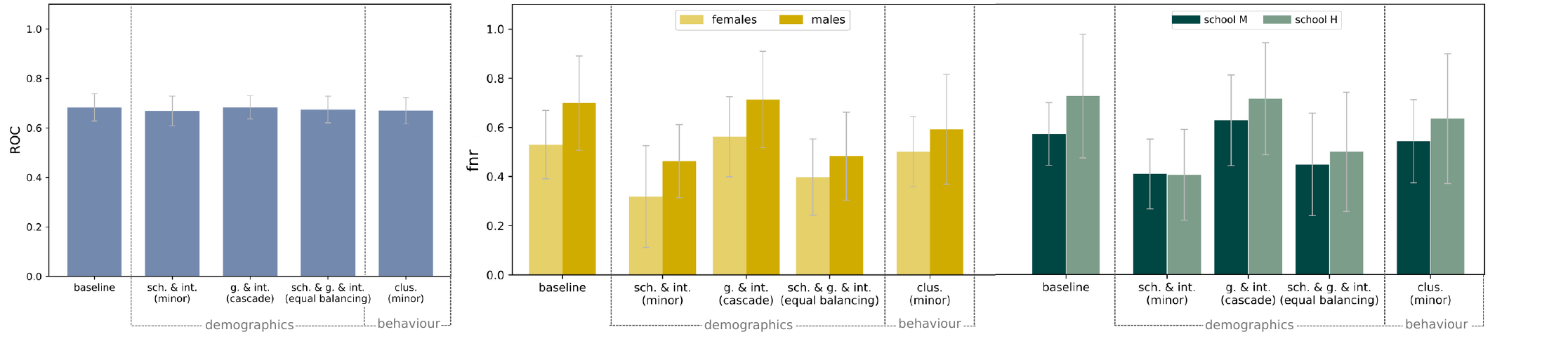}
  \caption{TugLet. Average and standard deviation over 10 folds of the false negative rate across genders for the baseline, the guided attribute oversampling and the behavioral oversampling. (left): auc over the entire dataset, (center): fnr across different genders, (right): fnr across different schools. Legend - sch.: school, int.: intervention, g.: gender, clus.:cluster.}\label{figure:tuglet:rocfnr}
  \Description{TugLet. Average and standard deviation over 10 folds of the false negative rate across genders for the baseline, the guided attribute oversampling and the behavioral oversampling. (left): auc over the entire dataset, (center): fnr across different genders, (right): fnr across different schools. Legend - sch.: school, int.: intervention, g.: gender, clus.:cluster. For Tuglet, clustering and school allied with gender and intervention reduced the fnr differences the most}
\end{figure}

To test our hypothesis, we added the combinations gender-intervention need, school-intervention need and gender-school-intervention need to our set $\mathcal{O}$, building the basis for our guided attribute upsampling process. Figure \ref{figure:tuglet:rocfnr} (left) illustrates the accuracy of the detector in terms of AUC for the baseline (no upsampling) and the different upsampling combinations in terms of upsampled attributes and sampling method used. We note that the upsampling does not significantly impact predictive performance with respect to the baseline. The fairness of the classifier in terms of FNR for the attribute gender for the baseline and the guided attribute upsampling combinations is displayed in Figure \ref{figure:tuglet:rocfnr} (middle). Similarly, we visualize the FNR for the different school districts obtained with and without upsampling in Figure  \ref{figure:tuglet:rocfnr} (right). Note that while we ran the experiments with all described upsampling techniques, we report only the result of the most successful upsampling technique as described in Section \ref{sec:biasmitigation}. We observe that upsampling based on the combination of school and intervention needs, reduces the FNR for both school districts and leads to an unbiased detector in terms of FNR (School M: $0.41$, School H: $0.41$). However, with this combination, only a small reduction in the FNR difference between genders is achieved as the FNR difference is reduced to $0.14$ (from $0.17$). When upsampling on the combination of gender and intervention need, we manage to again reduce the FNR difference between school districts (from $0.16$ to $0.09$), but only achieve slight improvements for gender (FNR difference reduced from $0.17$ to $0.15$). However, when we upsample on the full combination of attributes (gender - school - intervention need), we achieve a substantial gain in fairness: for school, the difference in FNR is reduced from $0.16$ to $0.05$ and for gender from $0.17$ to $0.08$. The original FPR differences for both genders and school areas were initially relatively low (0.01 and 0.06 respectively). In the former case, oversampling with a combination that involved \textit{gender} retained the same balance (+/- 0.01). In all attributes oversampling cases, FPR differences across school areas either decreased or remained the same. 

\vspace{1mm} \noindent \textit{Flipped Classroom}. For this learning context, we found bias in the detector for both investigated attributes in our first analysis. While the data set already had an imbalance with respect to gender (under-representation of females), it was balanced with respect to the country of diploma. Therefore, similar to TugLet, we investigated the interaction effects between the demographic attributes and the labels. We observed that combined attributes led to an imbalance (males-Country CF: $29\%$, males-Country CO: $36\%$, females-Country CF: $20\%$, females-Country CO: $15\%$), with females from country CO being strongly under-represented. We also found that students needing intervention were generally under-represented in the data set (label $1$; $42\%$). Furthermore, we observed an interaction effect between country and label: $62\%$ of the struggling students obtained their diploma in country CO. We therefore hypothesized that the bias in the detector stemmed from an interaction effect between gender and country and that we could therefore mitigate this bias by upsampling on the combination of those attributes (\textbf{\textit{H2-2}}).

\begin{figure}[t]
  \centering
  \includegraphics[width=\linewidth]{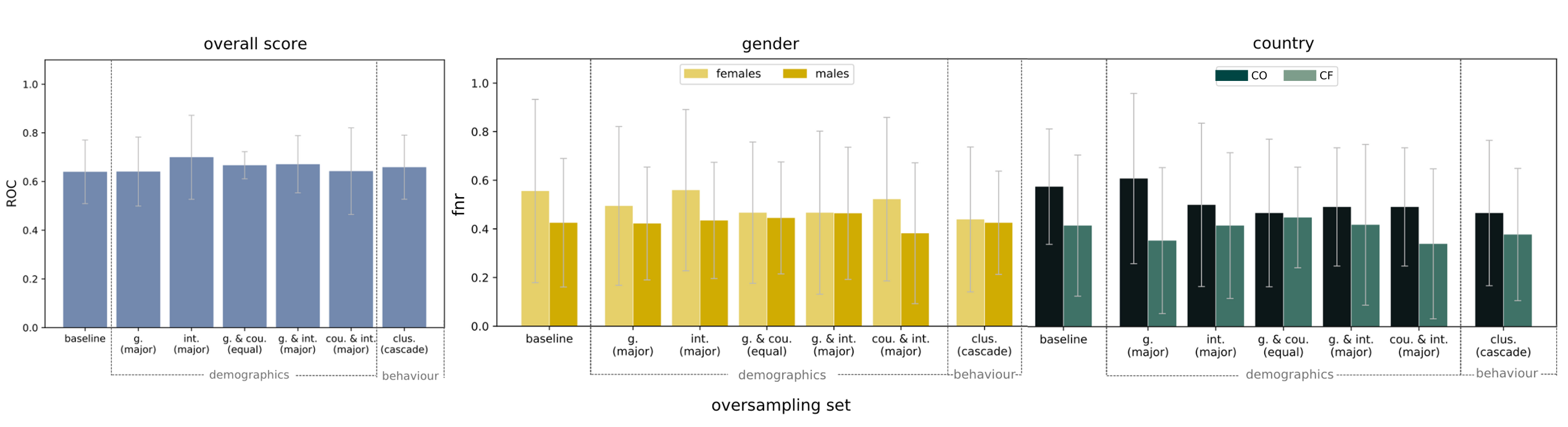}
  \caption{Flipped Classrooms. Average and standard deviation over $10$ folds of the false negative rate (fn) across genders for the baseline, the guided attribute oversampling, and the behavioral oversampling. Legend - cou.: country, int.: intervention, g.: gender, clus.:cluster}
  \label{figure:flipped:rocfnr}
  \Description{}
\end{figure}

Given our observations, we added the following combinations to our set $\mathcal{O}$ as a basis for the guided upsampling process: gender, gender-country, intervention need, gender-intervention need, country-intervention need and upsampled using the different techniques. The resulting AUC scores is reported in Figure \ref{figure:flipped:rocfnr} (left). We again display the baseline accuracy as well as the accuracy when upsampling on different combinations of attributes, reporting only the optimal upsampling technique for each combination. Upsampling leads to a gain in accuracy for all investigated attributes. This improved accuracy can be explained by the fact that the original dataset exhibited a class imbalance. We achieve the highest accuracy when oversampling on intervention need (AUC: $0.70$), but the difference in FNR is reduced only for country; it decreases from $0.16$ to $0.08$. The FNRs for gender and country achieved for the different upsampling methods are displayed in Figure \ref{figure:flipped:rocfnr} (middle + right). When upsampling on gender, as expected, the difference in FNR is reduced for gender (from $0.13$ to $0.07$), but the bias for country is even increased (from $0.16$ to $0.26$). Upsampling based on country and intervention need was unexpectedly not helpful, leading to only minor reductions in bias. By upsampling on country and gender, we managed to reduce the difference in FNR to $0.02$ for both attributes. In terms of FPR, a similar trend to what happened in \textit{Tuglet} was observed. Oversampling on demographics bore no improvement or was worse in terms of gender balance, but it reduced the differences across countries systematically. 

\vspace{1mm} \noindent \textit{In summary, the guided upsampling on combined attributes reduced detectors bias for both data sets. For both learning contexts, the largest reduction was achieved when oversampling on the combined demographic attributes, confirming both of our hypotheses \textbf{H2-1} and \textbf{H2-2}. We observed that for the flipped classroom data set, upsampling also improved the detector accuracy, probably due to the interaction effects between the label and demographic attributes. Moreover, the choice of the upsampling technique is essential; different techniques might be optimal based on the imbalance distribution in the data}.

\subsection{RQ3: Behavioral Oversampling}
\label{sec:rq3}
In our first two analyses, we observed that single demographic attributes are not always a reliable proxy for under-represented behaviors, but that combined demographic attributes seem to be. However, this might not be the case for all learning contexts and access to student demographics might be difficult. In our last analysis, we therefore investigated the effects of behavioral upsampling on both detector accuracy and fairness. 

\vspace{1mm} \noindent \textit{TugLet}. For TugLet, we used the six behavioral profiles identified and demonstrated to be associated with students' learning success by \cite{kaser2020modeling}. They are not balanced, i.e. the according clusters do not have the same size (systematic explorers: $14\%$, explorers: $23\%$, non explorers: $34\%$, slow explorers: $0.04\%$, mixed explorers: $0.13\%$, slow passers:$12\%$). Given the association between the detected behaviors and success and the under-representation of some type of behaviors, we hypothesized that upsampling on the identified profiles would decrease the bias of the detector (\textit{H3-1}).

To test this hypothesis, we upsampled based on the identified learning behaviors.
Figure \ref{figure:tuglet:rocfnr} (left) shows the accuracy in terms of AUC for the predictor after upsampling. While our experiments covered all upsampling techniques described in Section \ref{sec:biasmitigation}, we again report only the results of the most successful technique. Similar to the guided attribute oversampling, the behavioral upsampling does not lead to a significant performance drop. We then investigated the FNR of the detectors, as illustrated in Figure \ref{figure:tuglet:rocfnr} (middle + right). We observe that upsampling on behavior reduces the differences in FNR between genders from $0.17$ to $0.09$. For school, we achieve a reduction of the FNR from $0.16$ to $0.10$. From Figure \ref{figure:tuglet:rocfnr} (middle), we see that for gender, the reduction in bias is comparable to the decrease achieved by oversampling on combined attributes. For school, while the achieved decrease is substantial, it does not beat the reduction achieved by oversampling on combined attributes (see Figure \ref{figure:tuglet:rocfnr} (right)). On FPR rates, the differences across schools went from 0.06 to 0.02, and from 0.01 to 0.02 across genders.

\vspace{2mm} \noindent \textit{Flipped Classroom}. In the flipped classroom context, \cite{mejia2022identifying} clustered students based on their SRL behavior and demonstrated that the identified profiles were related to learning success. Again, the obtained clusters were not balanced (A: $15\%$, B: $16\%$, C: $26\%$, D: $24\%$, E: $20\%$). In a first analysis, it was found that cluster E  had a much lower rate ($21\%$) of required intervention compared to the overall data ($58\%$), while also having a substantially higher proportion of women than the general population. Given this imbalance across clusters and the interaction between SRL behavior and learning success, we again hypothesized that oversampling on the identified behavioral profiles would decrease the detector bias (\textit{H3-2}).

We therefore upsampled based on the identified behavioral profiles. Similar to the guided attribute oversampling, upsampling on behaviors led to an improvement in overall accuracy, as shown in Figure \ref{figure:flipped:rocfnr} (left). Moreover, by upsampling on behaviors, the difference in FNR between females and males was reduced from $0.13$ to $0.01$ (see Figure \ref{figure:flipped:rocfnr} (middle)). For the country of diploma, a decrease of unfairness from $0.16$ to $0.09$ was achieved (see Figure \ref{figure:flipped:rocfnr} (right)). For gender, the achieved reduction is in line with the decrease obtained via oversampling on combined demographics attributes. While the unfairness is almost halved also for country, the combined demographics oversampling was still more successful. FPR wise, though the difference across genders was significantly higher, it went from 0.05 to 0.01 across countries. 

\vspace{1mm} \noindent \textit{In summary, oversampling on behavior was successful in mitigating bias in the detectors for both learning contexts. While the behavioral oversampling did not outperform guided demographic oversampling, the obtained reductions in differences in FNR are comparable and we therefore accept \textbf{H3-1} and \textbf{H3-2}. Behavioral oversampling is therefore a valid approach for reducing bias when demographic attributes are not available.}

\section{Discussion}
\label{sec:discussion}
In this work we were broadly concerned with investigating and mitigating biases in models designed to early identify students at risk of failing.
We worked with datasets from a compulsory flipped-classroom course on linear algebra offered to university students and an open-ended exploration environment (TugLet) designed for middle-school students.
Although these educational contexts were very different, students in both were represented by both behavioral sequences of interaction and demographic attributes.
In both cases, the models' purpose was to detect at-risk students who might benefit from remedial sessions between their class/course and the final evaluation.

To answer our first research question, we investigated whether existing early-detection models trained on this data exhibited biases, and if these biases were related to demographic under-representation.
More specifically, we analyzed the performances of a bi-LSTM and an LSTM adapted to each dataset respectively, and we inspected the models' biases on gender and geographical attributes (country issuing the high school diploma for flipped classrooms and school district for TugLet).
We found differences in false-negative rates (FNR) across demographic groups in models trained on both datasets, which means that deployed models would exhibit biases against some demographic groups by not identifying them as needing help at higher rates.
In the case of the flipped-classroom model, these biases were aligned with under-representation in the dataset.
Women, students from School B, and students from Country 1 all had higher FNRs, and these groups were all underrepresented in the data.
However, in the case of the TugLet model, biases could not be connected to demographic under-representation.
In fact, TugLet data was almost perfectly balanced across gender and SES despite exhibiting biases against subgroups of these categories.

This finding helped motivate our second research question. We investigated whether oversampling on combined attributes (\textit{guided demographic oversampling}) might better mitigate model biases in comparison to single-attribute oversampling methods.
For both learning contexts, the answer was yes.
The largest reduction in bias was consistently achieved when oversampling on the combined demographic attributes.
Surprisingly, we also found that oversampling also improved the flipped-classroom model's overall accuracy.
We hypothesized that these gains in both bias reduction and accuracy were likely because the guided demographic oversampling approach made it possible to capture interaction effects between the label and different demographic attributes.
In other words, intersectional groups (e.g., females from a particular school) may exhibit specific behaviors related to educational outcomes, but that these specific behaviors are drowned out by other, more-dominant behavioral patterns when oversampling is only done on individual attributes.

While guided demographic oversampling was more effective than common demographic oversampling techniques, we identified a number of limitations with this approach.
First, this approach is not possible when extensive demographic data is not available (e.g., when only a single demographic attribute such as gender or country of origin is available).
Second, it is never possible to know for sure if the right combination of attributes has been found.
There is always the possibility that the model is biased against an intersectional group that has either not been upsampled, or for whom a specific demographic attribute was not collected.
For example, if the model is actually biased against Black women, but no information about race exists in the data, then it is not possible to use guided demographic oversampling to correct these biases.
Finally, even if detailed demographic data is available, organizing data into intersectional categories results in an exponential increase in the number of groups, which increases the threat of amplifying noise and complicates the choice of how to oversample the different groups.

We proposed a method to sidestep these issues called \textit{behavioral oversampling}.
This method was grounded in our hypothesis that demographic upsampling works because it forces the model to pay more attention to idiosyncratic behaviors exhibited by demographic groups who are underrepresented in the data.
So, rather than relying on the assumption that demographic attributes reliably serve as a proxy for behavior, why not directly oversample on behavior itself?
Our third and final research question was whether behavioral oversampling would also help to mitigate biases in our detectors, and if so how it would compare to the other demographic-based methods.
We first grouped students by clustering them in behavior space, and then upsampled under-represented groups to encourage the model to learn relationships between these behaviors and outcomes of interest.
This method consistently reduced biases across demographic groups, and these reductions were comparable to those obtained using demographic oversampling.

Though behavioral oversampling does manage to sidestep some of the limitations of demographic oversampling while simultaneously reducing model biases, it also has its share of limitations.
First, it relies on the ability to accurately cluster behavioral data into meaningful clusters.
An important direction for future work is to identify methods which produce meaningful clusters regardless of the data structure.
Second, when demographic information is unavailable, there is no way to confirm that this method has succeeded in reducing biases against specific demographic groups.
Still, the ability to mitigate biases between groups exhibiting similar behaviors is likely better than doing nothing at all.

Overall, our recommendations for those interested in mitigating algorithmic bias in the domain of early detection of at-risk students are as follows.
First, when multiple demographic attributes are available, we recommend examining model fairness within subgroups to account for potential intersectional biases.
Guided attribute oversampling can then be used in place of more common approaches to mitigate these biases.
Second, if students can also be clustered based on their behaviors, we recommend using behavioral oversampling as well.
In our experiments, behavioral oversampling worked as well as guided attribute oversampling and in general offers a privacy-friendly way to correct biases.

Finally, we have identified \textit{intersectional oversampling} as an important direction for future research.
Intersectional oversampling would proceed similarly to \textit{guided demographic oversampling} by grouping students based on combinations of demographic attributes and then oversampling these groups, but would differ in how bias is mitigated.
Instead of evaluating bias reduction within individual demographic groups, reduction in bias would be evaluated across intersectional groups.
This change would bring guided demographic oversampling more in-line with intersectional theory, and would continue the work started in this paper to make early detection models more fair and trustworthy.

\begin{acks}
This project was substantially co-financed by the Swiss State Secretariat
for Education, Research and Innovation SERI.
\end{acks}

\bibliographystyle{ACM-Reference-Format}
\bibliography{sample-base}


\end{document}